\documentclass[12pt,aps,showpacs,showkeys,preprintnumbers,amsmath,amssymb,nofootinbib,preprint]{revtex4}

\usepackage{adjustbox}
\usepackage{multirow}
\usepackage{subfigure}
\usepackage{graphicx}
\usepackage{color}
\usepackage{amsmath}
\usepackage{amsfonts}
\usepackage{amssymb}
\def \be  {\begin{equation}}
\def \ee  {\end{equation}}
\def \ee  {\end{equation}}
\def \bea {\begin{eqnarray}}
\def \eea {\end{eqnarray}}

\begin{document}

\title{Particle ratios with in Hadron Resonance Gas (HRG) and Artificial Neural Network (ANN) models}
\author{R. M. Abdel Rahman}
%\email{R.Mahmoud@eng.modern-academy.edu.eg}
\affiliation{Modern Academy for Enginnering and Technology, Basic Sciences Department, 11571, Mokattam, Cairo, Egypt}

\author{Mahmoud Y. El-Bakry}
%\email{Kemo_h3S@hotmail.com}
\affiliation{Ain Shams University, Faculty of Education, Physics Department, 11771, Roxy, Cairo, Egypt}

\author{D. M. Habashy}
%\email{Emad10@hotmail.com}
\affiliation{Ain Shams University, Faculty of Education, Physics Department, 11771, Roxy, Cairo, Egypt}

\author{Abdel Nasser Tawfik}
%\email{atawfik@nu.edu.eg}
\affiliation{Nile University - Egyptian Center for Theoretical Physics (ECTP), Juhayna Square off 26th-July-Corridor, 12588 Giza, Egypt}

\author{Mahmoud Hanafy}
\email{mahmoud.nasar@fsc.bu.edu.eg}
\affiliation{Physics Department, Faculty of Science, Benha University, 13518, Benha, Egypt}

\date{\today}

\begin{abstract}  	
Comparison between various particle ratios such as  $K^-/K^+$, $\pi^-/\pi^+$,  $\bar{p}/p$, $\bar{\Lambda}/\Lambda$, $\bar{\Sigma}/\Sigma$, $ \bar{\Omega}/\Omega$, $K^+/\pi^+$, $K^-/\pi^-$, $\bar{p}/\pi^-$, $p/\pi^-$, $\Lambda/\pi^-$, and $\Omega/\pi^-$ calculated using the HRG model with the results estimated from simulation and training of the ANN model in the presence of  different experiments measurements such as AGS, SPS, RHIC and LHC energies is done. The success of ANN simulation model to describe the results from both phonological (HRG) model and experimental data will encourage to use it in various predictions for other particle ratios in regions where is no experiments.
\end{abstract}

\keywords{Hadron Resonance Gas, Particles Ratios, ANN model}

\maketitle

%%%%%%%%%%%%%%%%%%%%%%%%%%%%%%%%%%%%%%%%%%%%%%%%%%%%%
%%%   Section I
%%%%%%%%%%%%%%%%%%%%%%%%%%%%%%%%%%%%%%%%%%%%%%%%%%%%%

\section{Introduction}

	Quark gluon Plasma (QGP) is a high temperature and density phase of powerfully interacting matter during which quarks and gluons are liberated from hadrons and are absolve to give the complete volume of the system \cite{cleymans1998thermal,Megias:2012hk}. The Quark-Gluon Plasma (QGP) and its phase transition on the Quantum Chromodynamics (QCD) phase diagram have been at the forefront of high energy physics research for the past few decades \cite{Vovchenko:2020lju,Lao:2018mxf}. In order to study the QGP and its thermodynamic behavior, many experiments have been undertaken to recreate this state of matter at particle colliders like the Relativistic Heavy-Ion collider (RHIC) at Brookhaven Laboratory  and the large hadron collider (LHC) at CERN, it's doable to recreate this state of matter and study its properties \cite{Tawfik:2014eba,tiwari2013particle,rafelski2000sudden,becattini2001features,tawfik2019equation}. So as for these accelerators to recreate conditions necessary to make this distinctive state of matter, terribly energetic beams of massive ions, like gold or lead nuclei, are collided head-on at speeds terribly close to the speed of light. The ensuing fireball, which may reach a temperature of many trillion degrees kelvin, is that the QGP \cite{cleymans1998thermal,Megias:2012hk,Vovchenko:2020lju,Lao:2018mxf,Tawfik:2014eba,tiwari2013particle,rafelski2000sudden,becattini2001features,tawfik2019equation}.

	In addition to experiment, several theoretical models are used to study the various signatures of QGP and have been verified through a comparison of the available experimental results. In this work, we are shown the particle ratios calculated using one of these models, the HRG model \cite{Hanafy:2020hgq,Hanafy:2021ffj,Ding:2021ajz,Du:2021zqz,Niida:2021wut,cleymans1993thermal,Vovchenko:2020lju,tawfik2016particle}.

	A thermal system of ideal, non-interacting particles that includes all experimentally known hadrons and their resonances. The idea has led to the formulation of the HRG model. During the last several decades, the ideal HRG model has shown remarkable agreement with both Lattice QCD and experimental results, up until the transition temperature \cite{Tawfik:2014dha,Tawfik:2013bza,Tawfik:2013pd,Tawfik:2012zp,bazavov2019chiral,shuryak2017strongly,Hagedorn:1965st}.

	Artificial neural networks are the modeling of the human mind with the simplest constructing blocks are known as neurons. Neurons are approximately one hundred billion inside the human brain \cite{PANG2019867,PANG2021121972,brouwer2014investigation}. In the human brain, facts is saved in this kind of manner as to be distributed, and we are able to extract a couple of piece of this facts while essential from our reminiscence in parallel. We aren't unsuitable while we are saying that a human brain is made of thousands of extreme effective parallel processors. In multi-layer artificial neural networks, there also are neurons located in a comparable way to the human brain. Each neuron is attached to different neurons with a certain coefficients. During training, information is shipped to those connection points with the result that the network is learned \cite{PANG2019867,PANG2021121972,brouwer2014investigation,Duarte:2020ngm,Mallick:2021wop,Li:2021plq,article3,Apolinario:2021olp,ELBAKRY2003995,Rojas:1996,article6,article7,DARWISH2015299}.\\

	Numerous a success applications of the Artificial Neural Networks (ANNs) in current years, have developed in nuclear and high-energy physics \cite{Habashy:2021qku,Habashy:2021orz,brouwer2014investigation,PANG2019867,PANG2021121972,Pandey_2016,Anil:2020lch,Duarte:2020ngm,Mallick:2021wop,Li:2021plq,article3,Apolinario:2021olp,ELBAKRY2003995}, in addition to in biology, chemistry, meteorology, and different fields of science. A most important intention of nuclear theory is to expect nuclear structure and nuclear reactions from many theories such as the theory of the strong interactions, Quantum Chromodynamics (QCD), Quark Gluon Plasma and extra theories. The particle ratio produced from a collision is one of the essential observables which has an extensive effect in the  particle production. Each set of the data refers to at least one event of the heavy-ion collisions \cite{Mallick:2021wop,Li:2021plq,Apolinario:2021olp}. Consequently, the overall performance of the ANNs taken into consideration on this work was evaluated using the calculations of the Chi-squared test \cite{Tawfik:2014dha,Haykin:2008,article5,cryst10040290,Rojas:1996,Stachel:1989pa,NADA201380}.\\

	The aim of this study is to compare between a well-known phenomenological model, the HRG model, and a powerful simulation model, ANN model, through computing various particle ratios produced from nuclear collisions as a function of center of mass energy in the presence of  different experimental data produced in high energy heavy-ion collision experiments \cite{afanasiev2002xi, alt2005omega, antinori2004energy,klay2003charged,chung2003near,pinkenburg2001production,afanasiev2002energy,klay2002longitudinal} including AGS, STAR, PHENIX, SPS, NA49, ALICE, and shall cover the future facilities NICA and FAIR \cite{Taranenko:2020vqn}.\\

	The present paper is organized as follows. Sec. \ref{models} gives short reviews on the two used models; HRG, and ANN models. In Sec. \ref{sec:Results}, the results of different particle ratios are presented. Sec. \ref{conc} where we discuss our results to achieve the aim of the present paper and to give the conclusion. 
	
\section{The used Models}
\label{models}

	In this section, both the computational numerical ANN and the phenomenological statistical HRG models will described here. The purpose of this paper is to train and simulate the ANN model in estimating various particle ratios generating from various experimental high-energy heavy-ion collisions through a comparison with a well-know phenomenological model; HRG model in the existence of the available various experimental data  \cite{tawfik2019equation,BraunMunzinger:2001mh}. The conclusion from this comparison shall come with a prediction of various particle ratios using the ANN model in regions where there are no experimental data yet.

\subsection{Artificial Neural Network method (ANN)}
\label{sec:ANN}
	Human brain is one of the complicated and the ultimate super machine which could carry out any complex mission with ease \cite{article1}. Artificial neural networks (ANNs) are the mathematical duplicate of brain structure that may imitative the manner brain works \cite{article2}. The essential precept of a neural network training is to modify the parameters to map the given connections among input and output \cite{inbook2}. It has been successfully used for solving various problems such as curve fitting, classification, pattern recongisiztion, system estimation, and plenty of more statistical problems \cite{article1,article2,inbook2}.\\
 
\begin{figure}[htbp]
\includegraphics[width=0.4\linewidth]{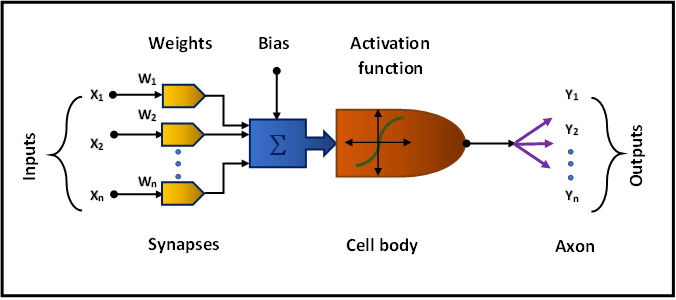}
\caption{The data processing of ANN simulation model for a neuron input signal to be processed and converted to the desired output signal.}
\label{fig:one}
\end{figure}

	ANN’s ability to solve such computational problem is based on the architecture of mathematical neurons that can reproduce part of the flexibility and power of human brain by learning from mistakes. Biological neurons have axons, dendrites and synapse which are modeled as interconnections, weights and activation function in mathematical neurons as given by McCulloch-Pitts model as shown in fig.(\ref{fig:one}). The output of neurons represents the activation function applied on the weighted sum of input signals and bias value \cite{Apolinario:2021olp,Duarte:2020ngm,Li:2021plq}.\\

	The neurons perform the weighted summation and non-linear mapping of input in a parallel manner \cite{HORNIK1989359}. In some cases, neurons are organized in sequential manner called layers, and are connected to both side of layers in-between input and output terminals. This type of architecture is called as multilayer artificial neural networks \cite{HORNIK1989359} and are more useful compared to single layer networks \cite{inbook1,Akkoyun:2019kve}.

The working of ANNs depends upon various interconnections defined as weights. In order to derive relation between both input and output data, these weights need to be adjusted properly. As the neural nets are derived from brain structure, they also are adapted in similar fashion human brain that learns different things. By learning from mistakes, human brain adjust their action. Similarly, neural network adjust their weights with respect to the difference between the expected outcome and the neural network output \cite{article5,article6}.

 Such steps by which the weights of neural networks are adjusted to fit the result are called as learning algorithms \cite{article5,article6}. More precisely, if the network has information about outcomes for a given set of input signal, they are characterized as supervised learning algorithm.
 
The error is back-propagated throughout the network similar to negative feedback in control system problem. The learning algorithms are nothing but numerical optimization techniques calculate parameters where the activation function has minimum value on the domain. The MSE value can be calculated as in Eq. (\ref{equ:one})\cite{Habashy:2021orz}:\\

\begin{equation}
MSE = \sum_{i=1}^{n} [\frac{(y_s-y_o)^2}{n}],   \label{equ:one}
\end{equation}

where $n$ is the number of datasets used for training the network, $y_s$ is the mean of the simulated result, and $y_o$ is the corresponding experimental value \cite{article5,Haykin:2008,article6}.\\

	Logarithmic sigmoid function (logsig) and a hyperbolic tangent sigmoid function (tansig) are the most activation function used in the hidden layers, their mathematical formula appear in Eqs.(\ref{equ:two}) and (\ref{equ:three}) respectively, whereas a pure linear function (purelin) is used in the output layer \cite{Anil:2020lch,Pandey_2016,inbook4}. 
	
\begin{equation}
(logsig)f(x)= \frac{e^x-e^{-x}}{e^x+e^{-x}},	\label{equ:two}	
\end{equation}

\begin{equation}
(tansig)f(x) = \frac{1}{1+e^{-x}},		\label{equ:three}
\end{equation}	
	
	A Purelin function is expressed as in Eq. (\ref{equ:four}) and it is defined as a neural transfer function that calculate a layer’s output from its net input and its formula.

\begin{equation}
A = purelin(N,FP).							\label{equ:four}	
\end{equation}

where $A$ is the output, $N$ is S-by-Q matrix of net input vectors, and $FP$ is struct of function parameters.\\

	While training the ANN model, different sets of internal network parameters are used by determining the number of hidden layer and number of neurons in both the hidden layer and transfer function.  Selection of the neural network architecture depends on the number of neurons in the hidden layer, and it represents an important issue as their choice varies from case to case basis. 

Using "squashing" functions or sigmoid transfer functions in the hidden layers is powerfully useful in calculations. They compress an infinite input range into a finite output range. Sigmoid functions are characterized by their slope which must tends to zero as the input gets large. This causes a problem when using steepest descent to train a multilayer network with sigmoid functions, which gives small changes in the weights and biases, even though the weights and biases are far from their optimal values.

	The present work uses a $Back Propagation ANN$ method and $Rprop$ training data to estimate different particle ratios such as $K^-/K^+$, $\pi^-/\pi^+$, $\bar{p}/p$, $\bar{\Lambda}/\Lambda$, $\bar{\Sigma}/\Sigma$, $ \bar{\Omega}/\Omega$, $K^+/\pi^+$, $K^-/\pi^-$, $\bar{p}/\pi^-$, $p/\pi^-$, $\Lambda/\pi^-$, and $\Omega/\pi^-$ produced from  heavy-ion collisions experiments at energies spanning from low energies to high energies including RHIC-BRAHMS, and  shall cover the future facilities NICA and FAIR \cite{Taranenko:2020vqn}.\\ 

	The algorithm $Resilient back-propagation algorithm$ (Rprop)is used. The trainrp function for training process is commonly used according to (Rprop) algorithm. Trainrp is a network training function that depends upon make updates for weights and biases. The (Rprop) algorithm introduced as first-order learning methods for neural networks. It considered one of the best performing ANN learning methods utilized \cite{PANG2019867,PANG2021121972}.

	The importance of using the $Resilient backpropagation$ (Rprop) training algorithm is to cancel the defect of the magnitudes of the partial derivatives of arbitrary error measure that is differentiable with respect to the weights. Only the sign of the derivative is used to determine the direction of the weight update; the magnitude of the derivative has no effect on the weight update. The size of the weight change is determined by a separate updated value \cite{ELBAKRY2003995,Rosenblatt:1958}.\\
	
\begin{figure}[htbp]
\includegraphics[width=0.4\linewidth]{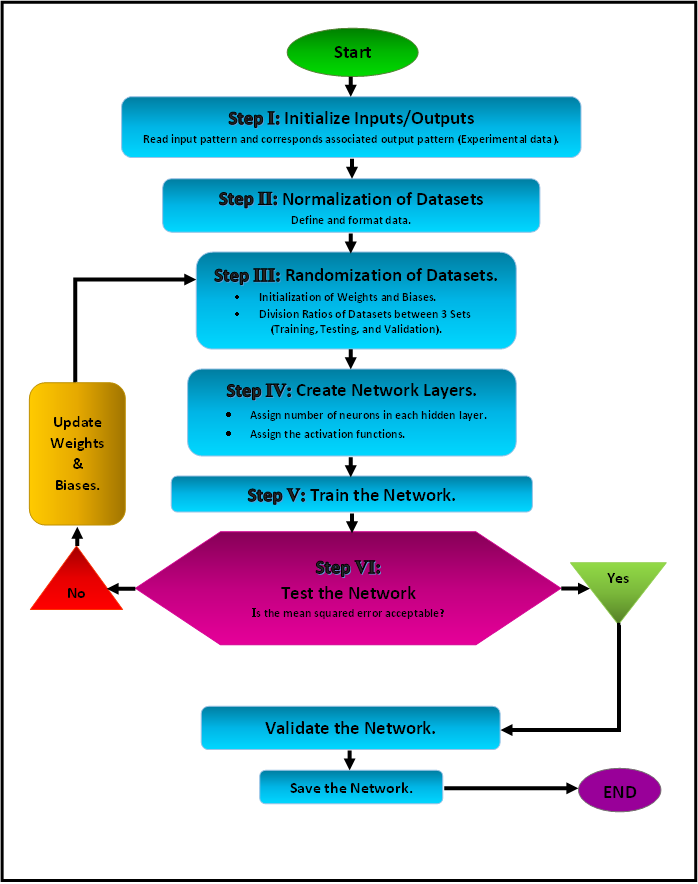}
\caption{Neural network architecture with weighted inputs and embedded transfer function.}
\label{fig:two}
\end{figure}

	In order to limit the calculations to a minimum time, certain limitations like number of iterations (epochs), gradient of convergence, and minimum mean square error (MSE), are used. The learning process is stopped if either one of the mentioned conditions are satisfied. The used algorithm is shown in a successful steps in fig. (\ref{fig:two}).\\

	Simulation and prediction are extended by using Matlab programming language interface for all considered particles ratios.  For a better estimations of the suggested particle ratios, the used hidden layers are tripled, where each one includes 10 neurons and the first starting applying epochs (refers to one cycle through the full training datasets)is 1000. The suitable activation function used for hidden layers is $logsig$ whereas the $purelin$ function used for the output layer.\\

\subsection{Hadron resonance gas (HRG) model} 
\label{sec:HRG}  

	The idea behind the HRG model was initially represented by R. Hagedorn by proposing two essential aspects, first one is about hadrons that they had an exponentially increasing mass spectrum, and  the second one about modeling the hadrons and their resonances as a non-interacting gas \cite{Tawfik:2016jzk,Megias:2012hk,Vovchenko:2020lju,Hagedorn:1965st,hagedorn1980hot}. 
	
	It can be understood in a bootstrap picture, i.e., the resonances are conjectured to be composed of further resonances, which in turn are consistent of lighter resonances and so on \cite{Tawfik:2016jzk,Megias:2012hk,Vovchenko:2020lju,Hagedorn:1965st,hagedorn1980hot}.
	 The HRG model as a theoretical model has been widely used to study phenomenologically the hadronic phase of QCD in comparison with lattice data, the well-known numerical approach \cite{satarov2009equation,Huovinen:2017ogf,karsch2005thermodynamic}.

	This model has been extremely effective in describing both the  production of heavy ion collision particles during chemical freezing and the computed lattice observables. The HRG model furnished a simple description of the degrees of freedom in the confined phase, and it is therefore relatively easy to carry out physical effects to test new physics and to income the constraints decreed by the experimental framework \cite{andronic2012interacting,Huovinen:2017ogf,samanta2019exploring}.

	The thermodynamic properties of such a system can be derived directly by statistical method which based upon the partition function definition which include all of the essential characteristics about the system \cite{andronic2012interacting,begun2013hadron,huang1963statistical}. For each hadron and its associated resonances, the individual particle partition function is $Z(T,\mu,V)$, and it has been expressed as a grand canonical ensemble as\cite{Hanafy:2021ffj,Tawfik:2019qrx}
	
\bea 
Z(T,\mu,V)=\mbox{Tr}\left[\exp\left(\frac{{\mu}N-H}{T}\right)\right], \label{equ:five}
\eea
where $V$ is the total volume of the system, $T$ is the temperature, $\mu$ is the chemical potential, $H$ is Hamiltonian, which combining a summation of the kinetic energies of the relativistic Fermi and Bose particles, counting degrees of freedom in deconfined and strongly interacting system, and $N$ is the number of constituents. 

	In the HRG model, Eq. (\ref{equ:five}) can be expressed as a sum over all hadron resonances \cite{Tawfik:2014eba,samanta2019exploring,Hanafy:2020hgq}. The hadron information employed in this calculations is taken from the recent particle data group \cite{vovchenko2017chemical,ParticleDataGroup:2018ovx}.\\

\begin{equation} 
\ln  Z(T,\mu,V)=\sum_i{{\ln Z}_i(T,\mu,V)} =\frac{V g_i}{2{\pi}^2}\int^{\infty}_0{\pm p^2 dp {\ln} {\left[1\pm {\lambda}_i \exp\left(\frac{-{\varepsilon}_i(p)}{T} \right) \right]}}, \label{equ:six}
\end{equation}

where $g_i$ are the individual degeneracies of each particle, $p$ is the momentum, $\pm$ stands for bosons and fermions, respectively, $\varepsilon_{i}=\left(p^{2}+m_{i}^{2}\right)^{1/2}$ is the dispersion relation of the $i$-th particle and $\lambda_i$ is its fugacity factor \cite{Tawfik:2014eba,Huovinen:2017ogf}.

\begin{equation} \label{equ:seven}
\lambda_{i} (T,\mu)=\exp\left(\frac{B_{i} \mu_{\mathtt{b}}+S_{i} \mu_{S}+Q_{i} \mu_{Q}}{T} \right),
\end{equation}

where $B_{i}$,$S_{i}$, and $Q_{i}$ are baryon, strange, and charge quantum numbers, while $(\mu_{\mathtt{b}})$, $(\mu_{S})$, and $\mu_{Q}$ are their corresponding chemical potentials of the $i$-th hadron, respectively. 

	The number density can be defined in the grand canonical ensemble as \cite{Tawfik:2014eba}:\\

\begin{equation} \label{equ:eight}
n_i(T,\mu)=\sum_{i}\frac{\partial\, {\ln Z}_i(T,V,\mu)}{\partial \mu_{i}}=\sum_{i}\frac{g_{i}}{2\pi^{2}}\int_{0}^{\infty }\frac{p^{2} dp}{\exp\left[\frac{{\mu}_{i} - {\varepsilon}_{i}(p)}{T}\right] \pm 1}.
\end{equation}

	The baryon chemical potential can be replaced in terms of the center of mass energy $\sqrt{S_{NN}}$ \cite{tawfik2015thermal}.\\  
 \begin{equation} \label{eq_9}
 \mu_{B}= \frac{a}{1+b \sqrt{S_{NN}}},
 \end{equation}
 
Where $a$, and $b$ are arbitrary constants, $a=1.245 \pm 0.094$ $GeV$, and $b=0.264 \pm 0.028$ $GeV^{-1}$ \cite{Hanafy:2020hgq}.

	Also, the temperature can also be defined in terms of the center of mass energy $\sqrt{S_{NN}}$ \cite{Tawfik:2016jzk}.
 \begin{equation}  \label{eq_10}
 T = T_{lim} \left[ \frac{1}{1+exp\left( \frac{1.172-log(\sqrt{S_{NN}}}{0.45})\right) }\right]. 
 \end{equation}
 
Where $\sqrt{S_{NN}}$ is taken in GeV, and $T_{lim} =  161\pm 4\hspace{0.05cm}MeV$.\\ 
    
   In this work, we have worked on hadrons and resonances including all decay channels with masses up to 12 GeV \cite{ParticleDataGroup:2018ovx}. All suggested particle ratios are calculated according to Eq. (\ref{equ:eight}) where the freezeout parameters are calculated using Eqs. (\ref{eq_9}) and (\ref{eq_10}).
 
\section{Results and Discussion}
\label{sec:Results}  

	The present study is focused on estimating various particle ratios such as  $K^-/K^+$, $\pi^-/\pi^+$,  $\bar{p}/p$, $\bar{\Lambda}/\Lambda$, $\bar{\Sigma}/\Sigma$, $ \bar{\Omega}/\Omega$, $K^+/\pi^+$, $K^-/\pi^-$, $\bar{p}/\pi^-$, $p/\pi^-$, $\Lambda/\pi^-$, and $\Omega/\pi^-$ measured at various energies from different heavy-ions collision experiments using one of the most powerful simulation tools, ANN model. The used mentioned experiments are the Alternating Gradient Synchrotron (AGS) \cite{PhysRevLett.88.102301,PhysRevC.62.024901,PhysRevC.60.064901,200053,PhysRevLett.90.102301,braun1999chemical}, STAR-RHIC \cite{STAR:2017sal,STAR:2009sxc,PhysRevC.79.034909,PhysRevLett.92.112301,Bzdak:2019pkr}, the Super proton Synchrotron (SPS) \cite{PhysRevC.73.044910,PhysRevC.69.024902}, NA49 \cite{NA49:2007stj, NA49:2002pzu,anticic2004lambda,afanasiev2002xi,alt2005omega}, NA57-SPS \cite{antinori2004energy}, PHENIX@RHIC \cite{Adler:2006xd}, E895-AGS \cite{Lisa:2005vw}, and ALICE \cite{Adam:2015kca}. The low energy region part of this range of the beam energy shall be accessed by NICA and FAIR future facilities \cite{Taranenko:2020vqn,tawfik2016particle}. The idea of this paper is to train and simulate ANN in predicting this various particles ratios and to compare with a phenomenological model such as HRG model.
	
The ANN model is perfectly trained based on the used experimental data, and all considered parameters are shown in Tab. (\ref{tab:2}). The ANN target is to get the best MSE value about ($10^{-6}$) according to Eq. (\ref{equ:one}). The considered particle ratios are calculated in the framework of HRG model according to Eq. (\ref{equ:eight}), where the dependence of the freezeout parameters on the $\sqrt{s}$, Eqs. (\ref{eq_9}) and (\ref{eq_10}), are summarized in Tab. (\ref{tab:template}).
\begin {table}[htbp] 
\caption{The dependence of the freeze-out parameters on the centre-of-mass energy.}
\centering
\begin{tabular}{|c|c|c|}
\hline 
 $\sqrt{{S}_{NN}} GeV$ &$\mu$ GeV & $ T$ GeV  \\ 
\hline 
 7 & 0.76 $\pm$ 0.005 & 0.053 $\pm$ 0.007  \\ 
\hline 
 7.7&0.7 $\pm$ 0.007 & 0.055 $\pm$ 0.005 \\
\hline 
 9&0.664 $\pm$ 0.006 & 0.061 $\pm$ 0.006  \\
\hline 
 11.5& 0.656 $\pm$ 0.004 &  0.073 $\pm$ 0.007\\ 
\hline 
 13 & 0.638 $\pm$ 0.007 & 0.084 $\pm$ 0.005 \\
\hline 
 19.6& 0.554 $\pm$ 0.007 & 0.091 $\pm$ 0.005 \\ 
\hline 
 27& 0.520 $\pm$ 0.005 & 0.102 $\pm$ 0.001 \\ 
\hline
 39& 0.470 $\pm$ 0.004 & 0.122 $\pm$ 0.002 \\
\hline 
 62.4& 0.404 $\pm$ 0.004 &0.138 $\pm$ 0.005 \\
\hline 
 130& 0.310 $\pm$ 0.003 & 0.145 $\pm$ 0.005\\
\hline 
 200& 0.263 $\pm$ 0.003 & 0.151 $\pm$ 0.006 \\
\hline 
\end{tabular}
\label{tab:template}
\end {table}

	Tab. (\ref{tab:2}) summaries conditions and parameters of our ANN for the four different particles.
 
\begin{table}[htbp]
\caption {All Conditions and parameters activated in current ANN for the considered particles ratios.}
\centering
\begin{adjustbox}{width=\columnwidth}
\begin{tabular}{|c|c|c|c|c|c|c|c|c|c|c|c|c|}  
\hline 
\multirow{2}{*}{ANN parameters} & \multicolumn{12}{|c|}{particles ratios}  \\
\cline { 2 - 13 }
  & $K^+/\pi^+$ & $K^-/\pi^-$  & $K^-/K^+$  & $\pi^-/\pi^+$  & $\bar{p}/p$  &  $\bar{\Lambda}/\Lambda$  &  $\bar{\Sigma}/\Sigma$ & $ \bar{\Omega}/\Omega$ & $\bar{p}/\pi^-$  & $p/\pi^-$  & $\Lambda/\pi^-$  & $\Omega/\pi^-$  \\ 
\hline 
 $Inputs$  & $\sqrt{{S}}$  & $\sqrt{{S}}$  & $\sqrt{{S}}$  & $\sqrt{{S}}$  & $\sqrt{{S}}$  & $\sqrt{{S}}$  & $\sqrt{{S}}$  & $\sqrt{{S}}$  & $\sqrt{{S}}$  & $\sqrt{{S}}$  & $\sqrt{{S}}$  & $\sqrt{{S}}$   \\
\hline
$Outputs$  & $K^+/\pi^+$  & $K^-/\pi^-$  & $K^-/K^+$  & $\pi^-/\pi^+$ &  $\bar{p}/p$  &  $\bar{\Lambda}/\Lambda$  &  $\bar{\Sigma}/\Sigma$ &  $ \bar{\Omega}/\Omega$  & $\bar{p}/\pi^-$  & $p/\pi^-$  & $\Lambda/\pi^-$  & $\Omega/\pi^-$    \\
\hline 
 Hidden layers &  3  & 3 & 3  & 3  & 3  & 3  & 3 & 3  & 3  & 3  & 3  & 3   \\
\hline 
 Neurons &  10,10,10 &  10,10,10 &  10,10,10 &  10,10,10  & 10,10,10 &  10,10,10  & 10,10,10  & 10,10,10 & 10,10,10  & 10,10,10  & 10,10,10 &  10,10,10 \\
\hline 
 Epochs &  1000  &  743 &  95 &  1000 &  31 &  90 &  1000 &  1000  & 15 &  1000 &  1000 &  79  \\
\hline 
 Training  algorithms&  Rprop   & Rprop  & Rprop  & Rprop  & Rprop &  Rprop &  Rprop &  Rprop &  Rprop &  Rprop &  Rprop &  Rprop   \\
\hline 
 Training functions& trainrp   & trainrp &  trainrp &  trainrp  & trainrp &  trainrp &  trainrp &  trainrp &  trainrp &  trainrp & trainrp &  trainrp   \\
\hline 
Transfer functions  & $logsig$   & $logsig$  & $logsig$ & $logsig$ &  $logsig$  & $logsig$ & $logsig$  & $logsig$ & $logsig$  & $logsig$  & $logsig$  & $logsig$   \\
\hline 
Performances $\times 10^{-6}$ &   $172$ &   $10$  &   $126$    &  $425$    & $9.8$  & $9.19$ &  $68.1$ &  $368$  & $2.56$  & $21.1$  & $147$ &  $9.46$   \\
\hline 
Output functions  & $ Purelin $  & $ Purelin $  & $ Purelin $  & $ Purelin $ &  $ Purelin $  & $ Purelin $ &  $ Purelin $ &  $ Purelin $ &  $ Purelin $ &  $ Purelin $ &  $ Purelin $ &  $ Purelin $  \\
\hline
 \end{tabular}
 \end{adjustbox}
 \label{tab:2}
\end {table}

The obtained results are classified into three groups:

\begin{itemize}
\item Anti-particle to particle versus $\sqrt{s}$ \\

In this part, the obtained training results for antiparticles to particle ratios, i.e., (a)$K^-/K^+$, (b)$\pi^-/\pi^+$, (c) $\bar{p}/p$, (d) $\Lambda/\Lambda$,(e) $\bar{\Sigma}/\Sigma$, and (f) $\bar{\Omega}/\Omega$, are shown in Figs. (\ref{fig:four} (a-f)). By training ANN, the network's mean squared error value is reduced from a huge value to a lower value, as shown from Figs. (\ref{fig:four} (a-f)) where the network is evolving. After the network has learned the training set, i.e., input values, the model training is finalized. Finally, the validation of the obtained results are takes place according to the best fit. The optimal ANN simulation and prediction model training, for these ratios is chosen as a result of MSE value of $126\times 10^{-6}$, $425\times 10^{-6}$, $9.8 \times 10^{-6}$, $9.19 \times 10^{-6}$, $68.1 \times 10^{-6}$, and $368 \times 10^{-6}$, respectively.\\

\begin{figure}[htbp]
\includegraphics[width=0.25\linewidth]{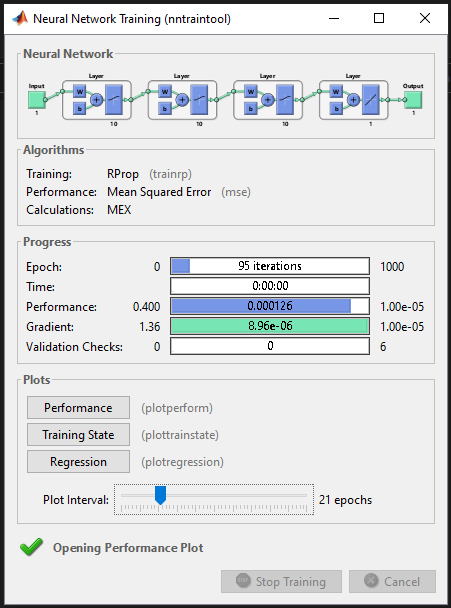}
\includegraphics[width=0.25\linewidth]{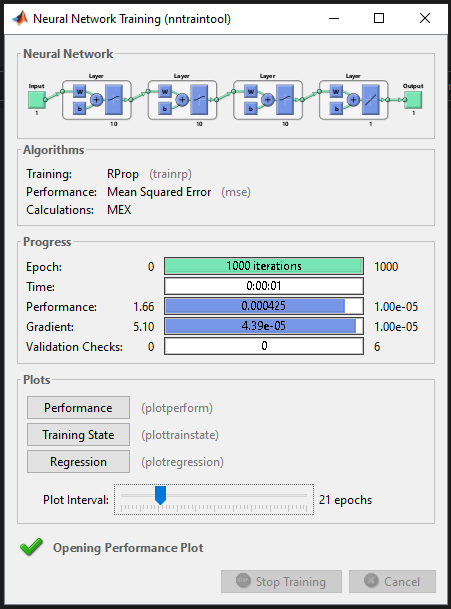}
\includegraphics[width=0.25\linewidth]{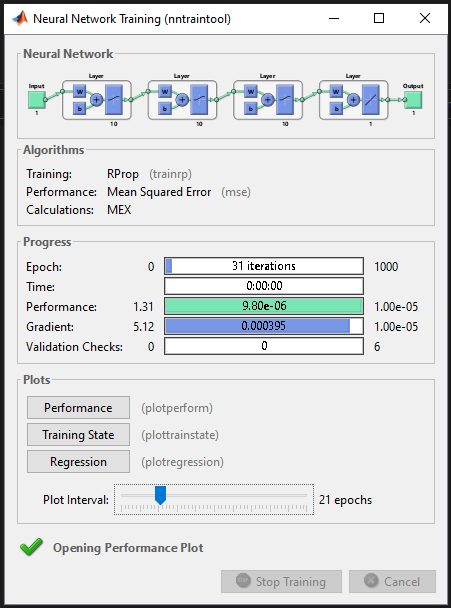} 
\includegraphics[width=0.25\linewidth]{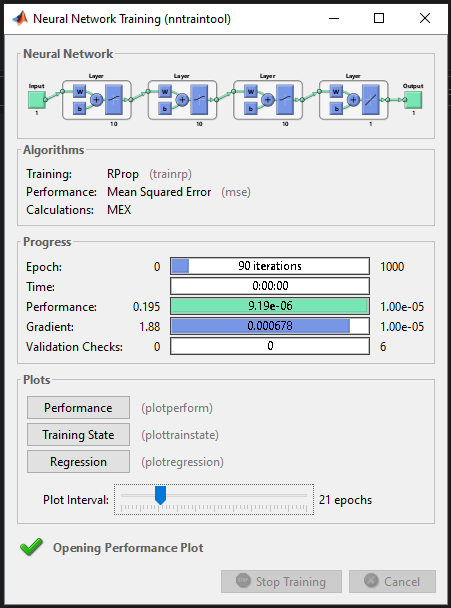}
\includegraphics[width=0.25\linewidth]{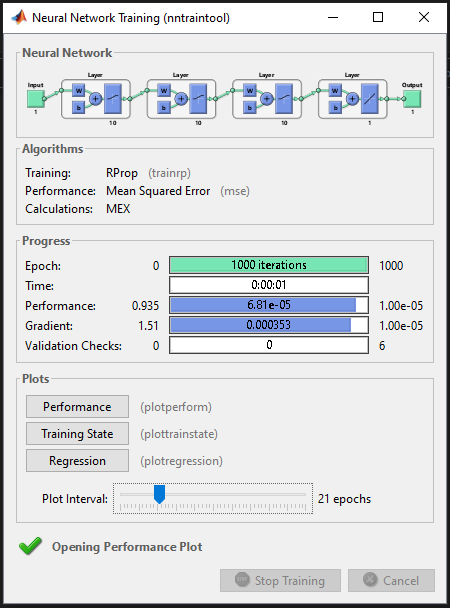}
\includegraphics[width=0.25\linewidth]{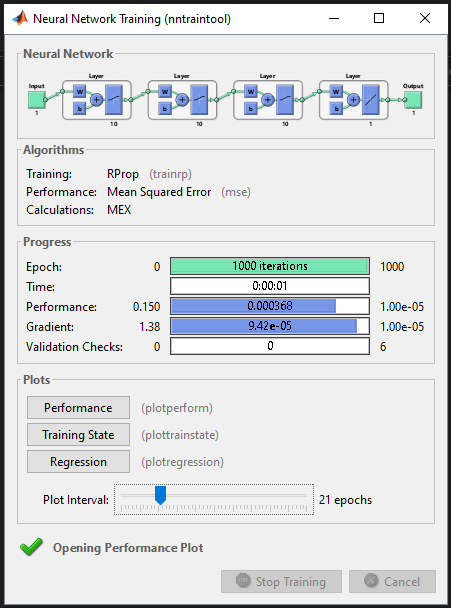}
\caption{Matlab neural network training tool used for estimating  particles ratios (a)$K^-/K^+$, (b)$\pi^-/\pi^+$, (c) $\bar{p}/p$, (d) $\Lambda/\Lambda$, (e) $\bar{\Sigma}/\Sigma$, and (f) $\bar{\Omega}/\Omega$, w.r.t. $\sqrt{{s}}$.}
\label{fig:four}
\end{figure}

The performance of the mentioned ratios are shown in Figs. (\ref{fig:five} (a-f)).

\begin{figure}[htbp]
\includegraphics[width=0.32\linewidth]{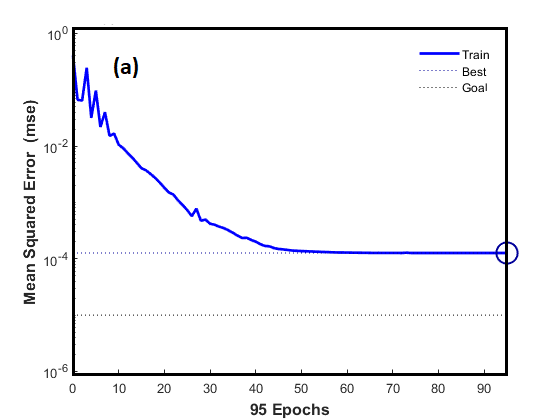} 
\includegraphics[width=0.32\linewidth]{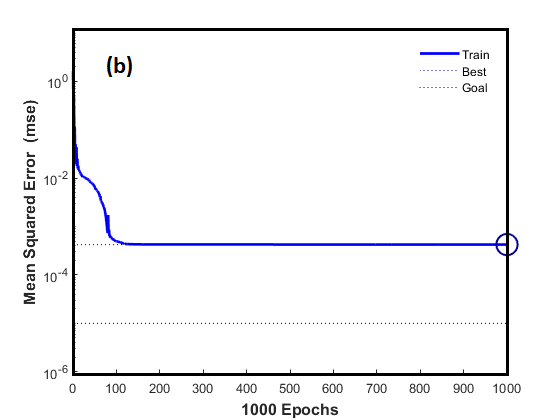} 
\includegraphics[width=0.32\linewidth]{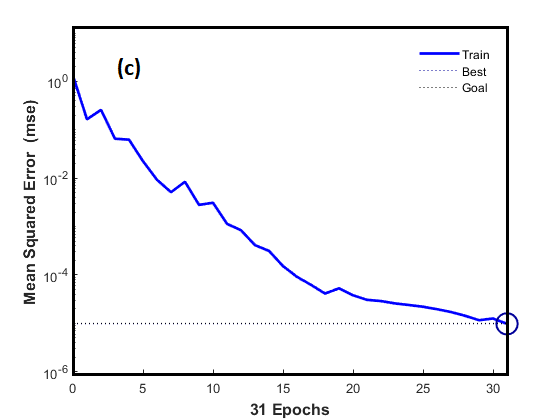} 
\includegraphics[width=0.32\linewidth]{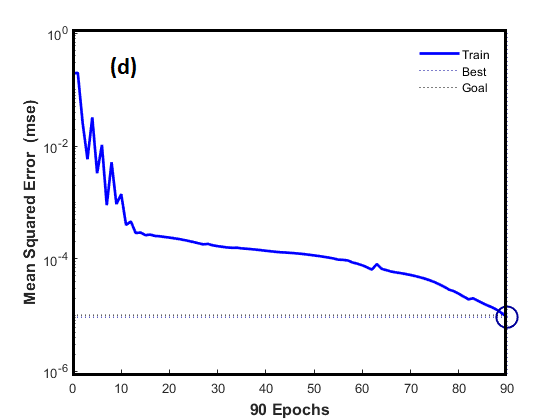} 
\includegraphics[width=0.32\linewidth]{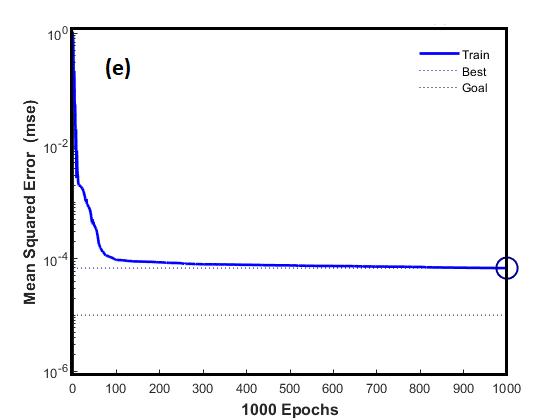} 
\includegraphics[width=0.32\linewidth]{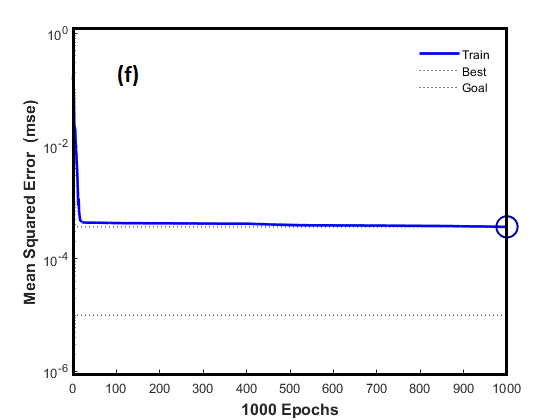} 
\caption{The performance of the used neural network for particles ratios (a) $K^-/K^+$, $\pi^-/\pi^+$, (b)$\bar{p}/p$,(c) $\bar{\Lambda}/\Lambda$, (d)$\bar{\Sigma}/\Sigma$, and (e) $ \bar{\Omega}/\Omega$ as a function of $\sqrt{s}$.}
\label{fig:five}
\end{figure}

Fig. (\ref{fig:three}) shows the antiparticle-to-particle ratios, (a)$K^-/K^+$, (b)$\pi^-/\pi^+$, (c) $\bar{p}/p$, (d) $\Lambda/\Lambda$, (e) $\bar{\Sigma}/\Sigma$, and (f) $\bar{\Omega}/\Omega$, estimated from ANN model (naive dashed lines) in comparison with that measured from the mentioned different experiments (symbols) \cite{afanasiev2002xi,alt2005omega,antinori2004energy,klay2003charged,chung2003near, pinkenburg2001production, afanasiev2002energy, klay2002longitudinal} and HRG calculations (black solid curves) \cite{Tawfik:2014eba,braun1999chemical,cleymans1993thermal,tiwari2013particle,rafelski2000sudden, becattini2001features,cleymans1998thermal}.

\begin{figure}[htbp]
\includegraphics[width=0.32\linewidth]{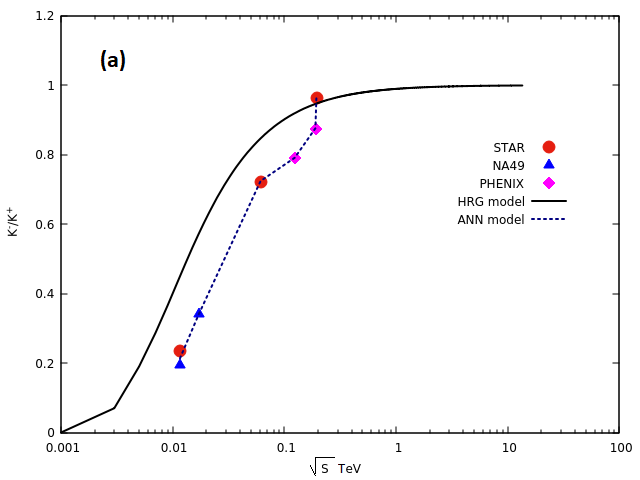}
\includegraphics[width=0.32\linewidth]{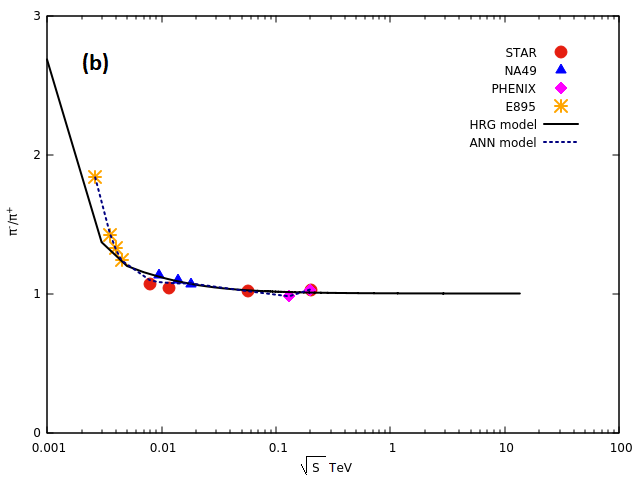}
\includegraphics[width=0.32\linewidth]{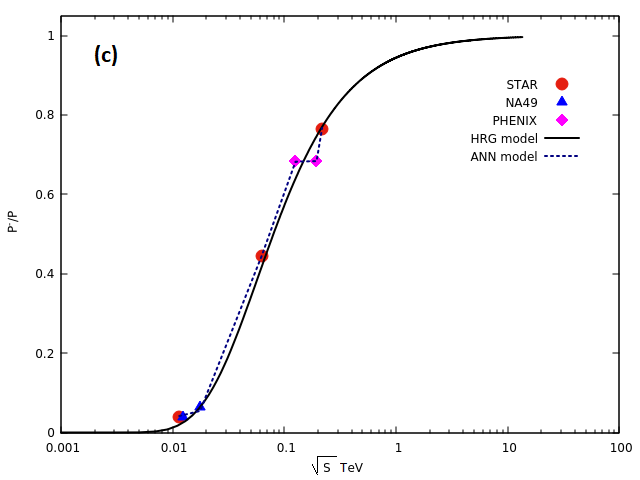}
\includegraphics[width=0.32\linewidth]{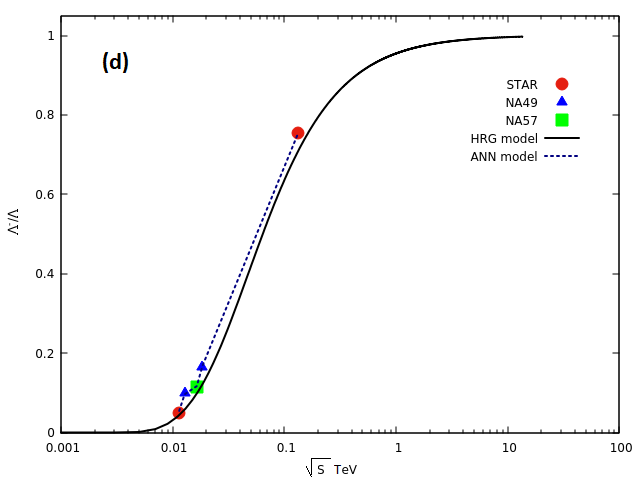}
\includegraphics[width=0.32\linewidth]{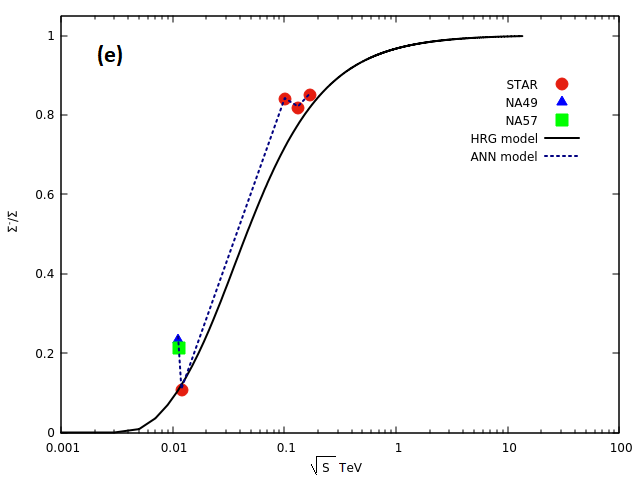}
\includegraphics[width=0.32\linewidth]{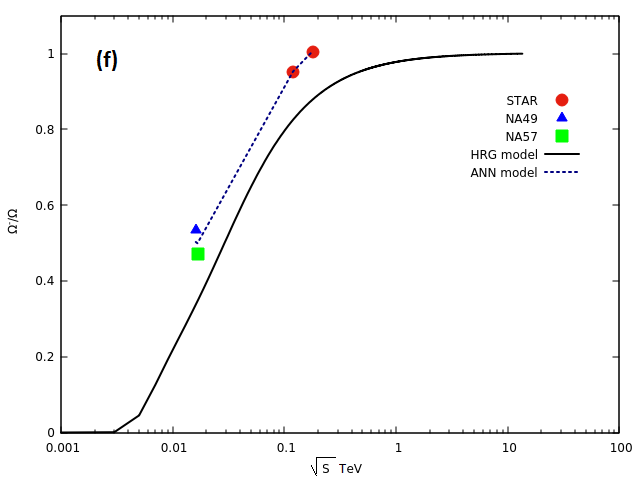}
\caption{Variation of particle ratios (antiparticle-to-particle pairs) measured from different experiments (symbols):(a)$K^-/K^+$, (b)$\pi^-/\pi^+$,(c) $\bar{p}/p$, (d) $\Lambda/\Lambda$,(e) $\bar{\Sigma}/\Sigma$, and (f) $\bar{\Omega}/\Omega$ \cite{Adler:2006xd,Adam:2015kca,afanasiev2002xi,alt2005omega,anticic2004lambda,antinori2004energy,klay2003charged,blume2005review,chung2003near,Lisa:2005vw,pinkenburg2001production,afanasiev2002energy,klay2002longitudinal}. The experimental data are compared to a numerical simulations by using ANN method (dashed lines), and also HRG calculations (solid curves) \cite{Vovchenko:2020lju,Tawfik:2014eba,braun1999chemical,tiwari2013particle,rafelski2000sudden, becattini2001features,cleymans1998thermal}.}
\label{fig:three}
\end{figure}

It is clear from Fig. (\ref{fig:three}) that ANN model has succeeded to predict the antiparticle-to-particle ratios as (a)$K^-/K^+$, (b)$\pi^-/\pi^+$, (c) $\bar{p}/p$, (d) $\Lambda/\Lambda$, (e) $\bar{\Sigma}/\Sigma$, and (f) $\bar{\Omega}/\Omega$ at all considered range of energy. The predictions of ANN model for the mentioned antiparticle-to-particle ratios agree well with the used experimental measurements than the HRG model especially for the ratios $\bar{\Sigma}/\Sigma$ and $\bar{\Omega}/\Omega$.	
	
\end{itemize}

\begin{itemize}

\item strange to non-strange mesons ratios versus $\sqrt{s}$ \\

The obtained training results for strange meson to non-strange meson ratios (a) $K^+/\pi^+$, and (b) $K^-/\pi^-$ \cite{afanasiev2002xi, alt2005omega, antinori2004energy} are shown in Fig. (\ref{fig:seven}(a-b)). By training the ANN model, we have obtained the optimum value for the network's mean squared error, as can be shown from the figure. The optimal ANN simulation model training, for theses particles ratios, $K^+/\pi^+$ and $K^-/\pi^-$, is chosen as a result of the lowest MSE value of $172\times 10^{-6}$, and $10 \times 10^{-6}$, respectively.	

\begin{figure}[htbp]
\includegraphics[width=0.25\linewidth]{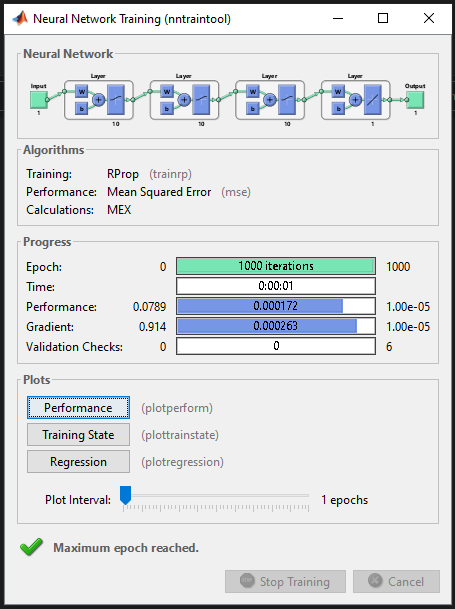} 
\includegraphics[width=0.25\linewidth]{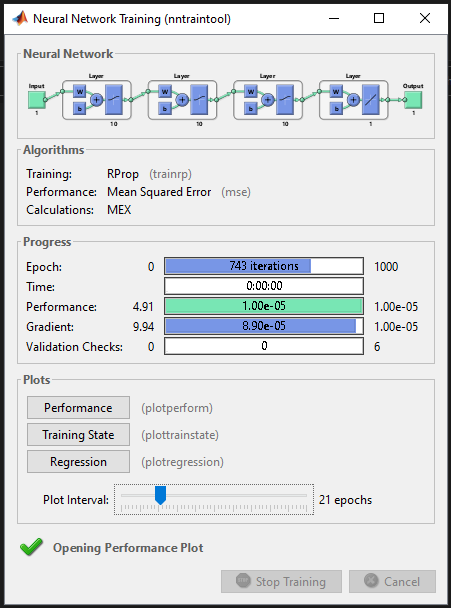}
\caption{Matlab neural network training tool used for estimating  strange to non-strange mesons ratios , $K^+/\pi^+$ and $K^-/\pi^-$, w.r.t. center of mass energy $\sqrt{{S}}$.}
\label{fig:seven}
\end{figure}

the lowest MSE value has reflected the good performance of the ANN training for these ratios as shown in Fig. (\ref{fig:eight} (a-b)).

\begin{figure}[htbp]
\includegraphics[width=0.32\linewidth]{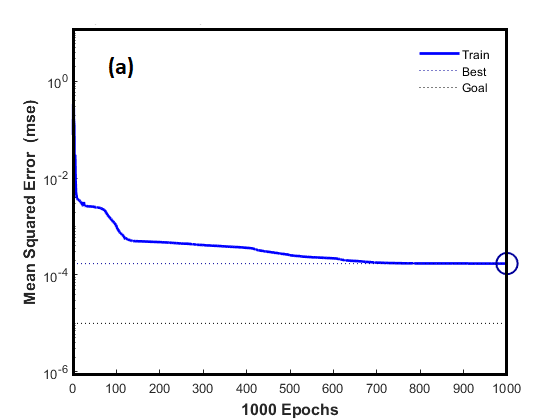} 
\includegraphics[width=0.32\linewidth]{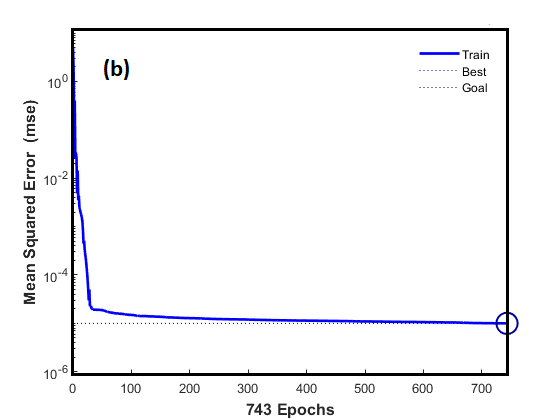} 
\caption{The performance of the used neural network for particles ratios (a) $K^+/\pi^+$, and (b) $K^-/\pi^-$ as a function of center of mass energy $\sqrt{{S}}$.}
\label{fig:eight}
\end{figure}
	Fig. (\ref{fig:six}) shows the same as in Fig. (\ref{fig:three}) but here for strange meson to non-strange meson pairs. Here, We are showing the estimated results from ANN model for some strange-to-non-strange ratios (naive dashed line) such as (a) $K^+/\pi^+$ and (b) $K^-/\pi^-$ in comparison with both the experimental data (symbols) \cite{afanasiev2002xi, alt2005omega, antinori2004energy} and the HRG model calculations (black solid curves). 
	
\begin{figure}[htbp]
\includegraphics[width=0.32\linewidth]{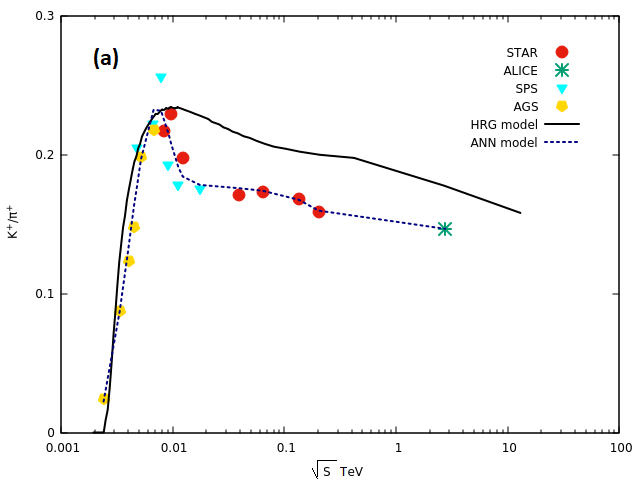}
\includegraphics[width=0.32\linewidth]{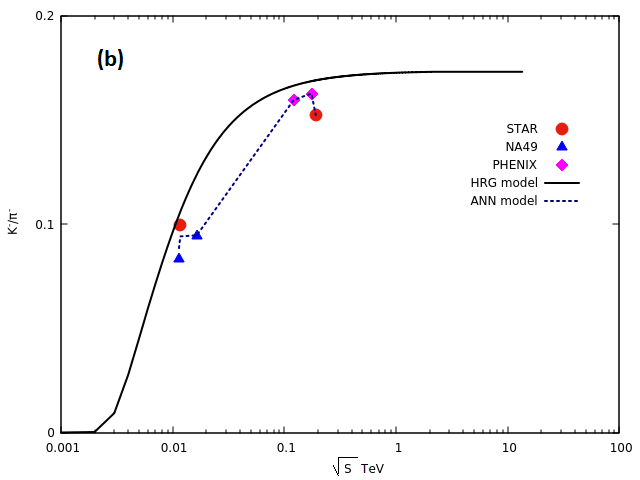}
\caption{The particle ratios (strange to non-strange mesons) measured from different experiments (symbols): (a) $K^+/\pi^+$, and (b) $K^-/\pi^-$ \cite{afanasiev2002xi, alt2005omega, antinori2004energy} are compared to a numerical simulations by using ANN method (naive dashed lines), and also HRG calculations (black solid curves).}
\label{fig:six}
\end{figure}	
As it clear from Fig. (\ref{fig:six}), the ANN model results for the particle ratios $K^+/\pi^+$, $K^-/\pi^-$ are agreed well with the available experimental measurements \cite{afanasiev2002xi, alt2005omega, antinori2004energy} and with that are calculated from the HRG model. 
	
\end{itemize}

\begin{itemize}

\item Baryon to meson ratios versus $\sqrt{s}$ \\
Fig. (\ref{fig:nine} (a-d)) shows the obtained training results for some baryon to meson ratios such as (a)$\bar{p}/\pi^-$, (b) $p/\pi^-$, (c) $\Lambda/\pi^-$, and (d) $\Omega/\pi^-$. The network's mean squared error value reduced from a huge value to a lower value after training the ANN model, as can be shown from the figure. The optimal ANN simulation model training, for the particles ratios $K^+/\pi^+$, and $K^-/\pi^-$, is chosen as a result of MSE value of $2.56\times 10^{-6}$, $21.1\times 10^{-6}$, $147\times 10^{-6}$, and $9.46 \times 10^{-6}$, respectively.	

\begin{figure}[htbp]
\includegraphics[width=0.24\linewidth]{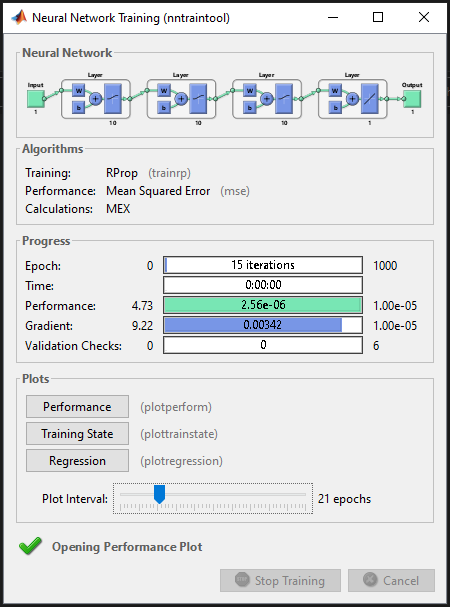} 
\includegraphics[width=0.24\linewidth]{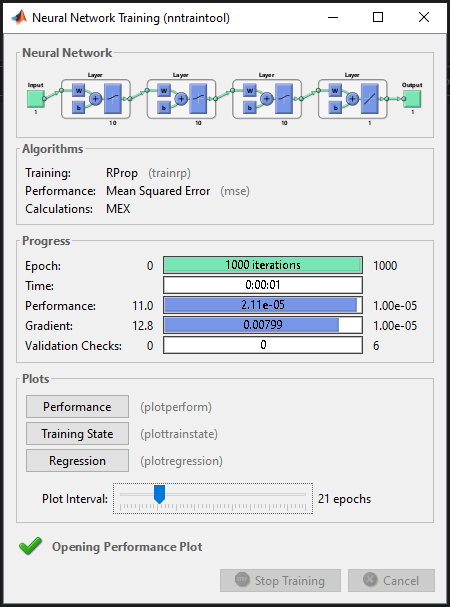}
\includegraphics[width=0.24\linewidth]{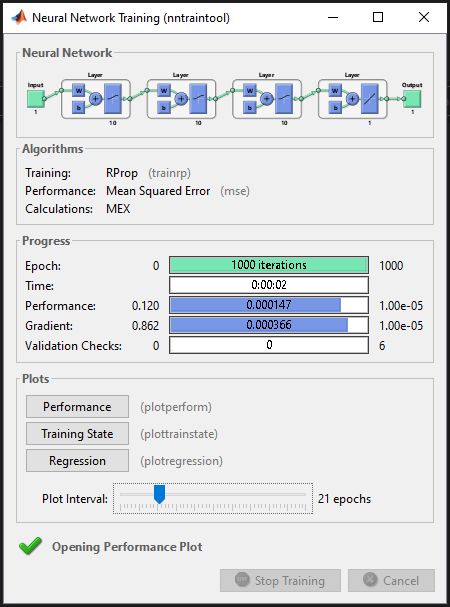}
\includegraphics[width=0.24\linewidth]{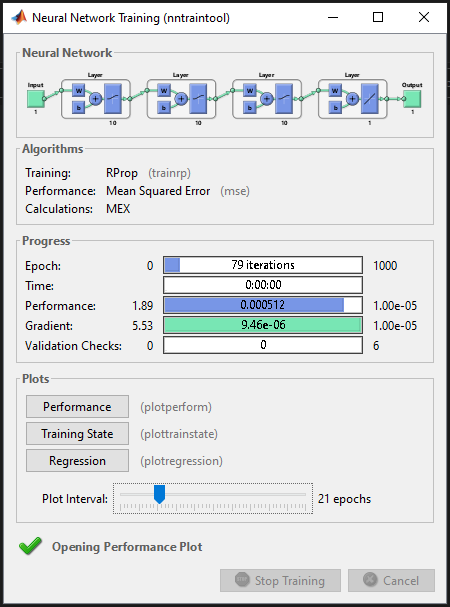}
\caption{Matlab neural network training tool used for estimating some baryon to meson ratios such as (a)$\bar{p}/\pi^-$, (b) $p/\pi^-$, (c) $\Lambda/\pi^-$, and (d) $\Omega/\pi^-$ w.r.t. center of mass energy $\sqrt{{S}}$.}
\label{fig:ten}
\end{figure}

The performance of the ANN model for these considered particle ratios are shown in fig. (\ref{fig:eleven} (a-d)).

\begin{figure}[htbp]
\includegraphics[width=0.32\linewidth]{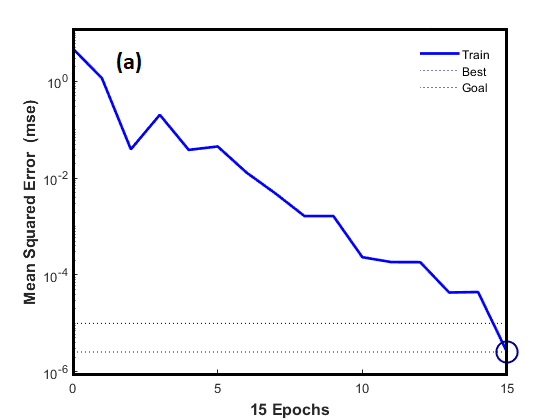} 
\includegraphics[width=0.32\linewidth]{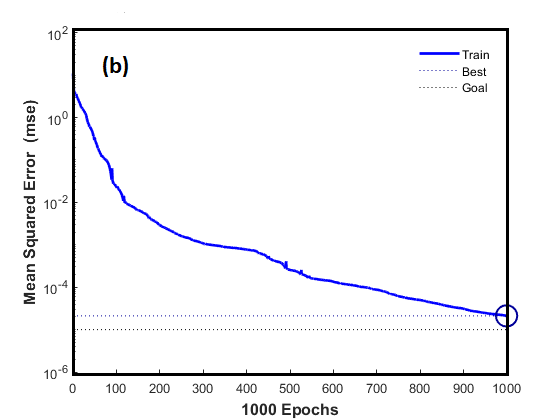} 
\includegraphics[width=0.32\linewidth]{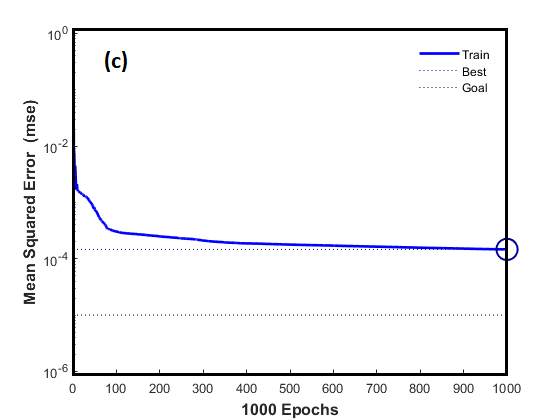} 
\includegraphics[width=0.32\linewidth]{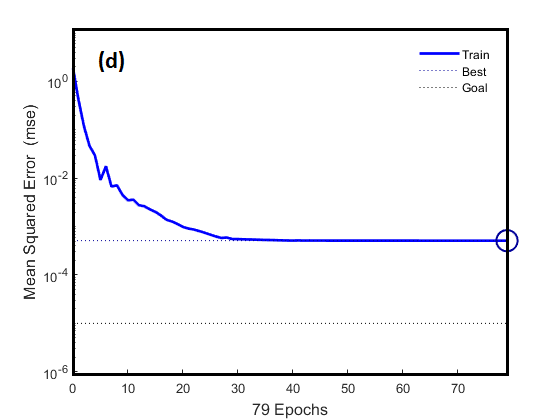}
\caption{The performance of the used neural network for particles ratios (a) $\bar{p}/\pi^-$, (b)$p/\pi^-$, (c)$\Lambda/\pi^-$, and (d)$\Omega/\pi^-$ as a function of center of mass energy $\sqrt{{S}}$.}
\label{fig:eleven}
\end{figure}
	Fig. (\ref{fig:nine}) shows the same as in figs. (\ref{fig:three}) and (\ref{fig:six}) but for antiparticle-to-particle pairs such as (a)$\bar{p}/\pi^-$, (b) $p/\pi^-$, (c) $\Lambda/\pi^-$ ,and (d) $\Omega/\pi^-$. The obtained results using ANN model is tackled against the available experimental data (symbols) \cite{afanasiev2002xi,alt2005omega,antinori2004energy,klay2003charged,chung2003near, pinkenburg2001production, afanasiev2002energy, klay2002longitudinal} and to that calculated from HRG calculations (black solid curves).\\

\begin{figure}[htbp]
\includegraphics[width=0.33\linewidth]{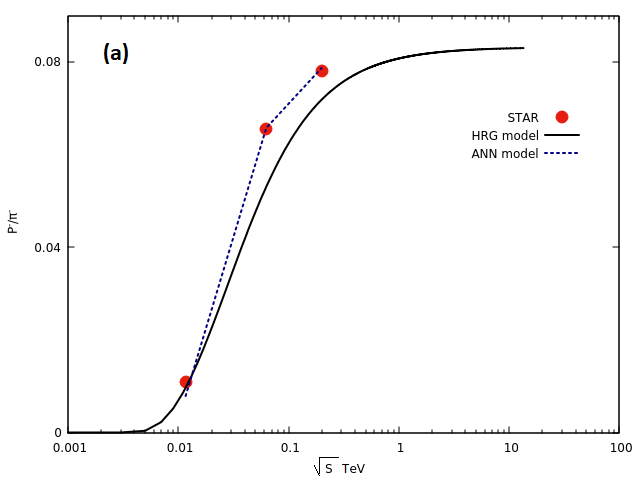}
\includegraphics[width=0.33\linewidth]{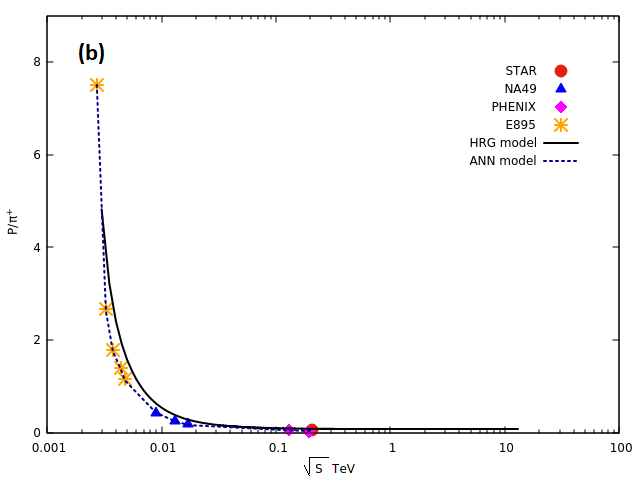}
\includegraphics[width=0.33\linewidth]{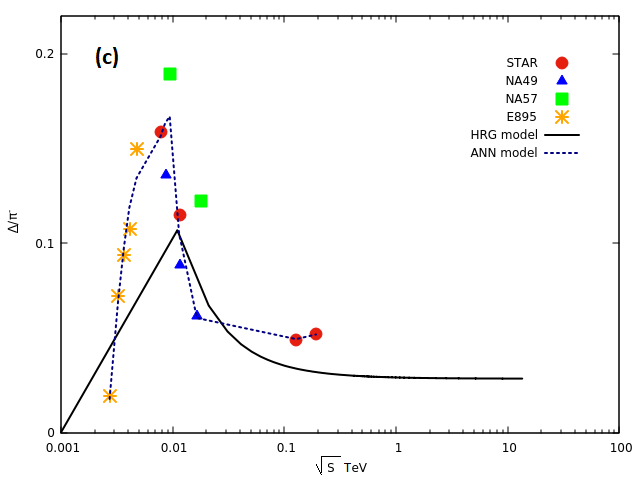}
\includegraphics[width=0.33\linewidth]{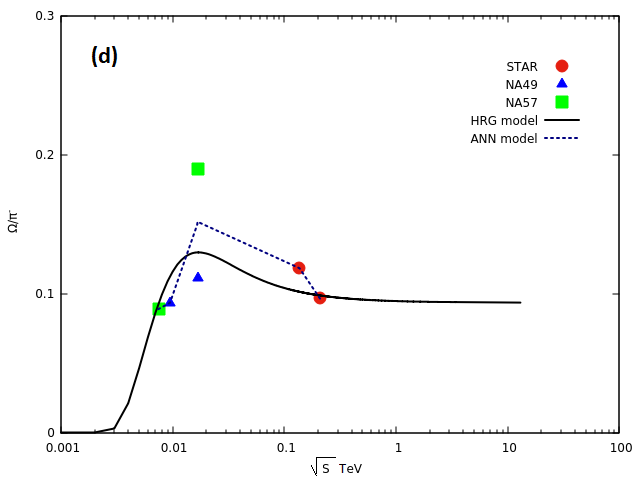}
\caption{Other particle ratios (baryon to meson pairs) measured from different experiments (symbols):(a)$\bar{p}/\pi^-$, (b) $p/\pi^-$, (c) $\Lambda/\pi^-$ ,and (d) $\Omega/\pi^-$ \cite{afanasiev2002xi, alt2005omega, antinori2004energy, klay2002longitudinal, chung2003near, pinkenburg2001production, afanasiev2002energy, back2004production}. The experimental data are compared to a numerical simulations by using ANN method (naive dashed lines),and also HRG calculations (black solid curves) \cite{Tawfik:2014eba, braun1999chemical, cleymans1993thermal, tiwari2013particle, rafelski2000sudden, becattini2001features,cleymans1998thermal}.}
\label{fig:nine}
\end{figure}

The particle ratios (baryon to meson pairs) computed from ANN model shows a good agreement with both the experimental measurements and the HRG model calculations. 
\end{itemize}
All discussed deduced particle ratios using ANN model prove the validity of the model to predict particle ratios resulted from high-energy heavy-ion collisions.
  
ALL particle ratios "$R$" can be estimated from ANN model as

\begin{equation}\label{equ:eleven}
\begin{array}{l}
R =\text { purelin }[\text { net. } L W\{4,3\} \text { logsig }(\text { net. } L W\{3,2\} \\
\text { logsig }(\text { net. } L W\{2,1\} \text { logsig }(\text { net.IW }\{1,1\} R+\text { net.b }\{1\}) \\
\quad+\text { net.b }\{2\})+\text { net.b }\{3\})+\text { net.b }\{4\}]
\end{array}
\end{equation}

Where $R$ is the input ( center of mass energy ),\newline
$IW$ and $LW$ are the linked weights as follow: \newline
$net.IW \left\{1, 1\right\}$ is linked weights between the input layer and first hidden layer,\newline
$net.LW \left\{2, 1\right\}$ is linked weights between first and second hidden layer,\newline
$net.LW \left\{3, 2\right\}$ is linked weights between the second and third hidden layer,\newline
$net.LW \left\{4, 3\right\}$ is linked weights between the third and output hidden layer,\newline
and $b$ is the bias and considers as follow\newline
$net.b\left\{1\right\}$ is the bias of the first hidden layer,\newline
$net.b\left\{2\right\}$ is the bias of the second hidden layer,\newline
$net.b\left\{3\right\}$ is the bias of the third hidden layer,\newline
$net.b\left\{4\right\}$ is the bias of the output layer.

\section{Conclusions}
\label{conc}

In this paper, the ANN simulation model is used for computing different particle ratios such as $K^-/K^+$, $\pi^-/\pi^+$, $\bar{p}/p$, $\bar{\Lambda}/\Lambda$, $\bar{\Sigma}/\Sigma$, $ \bar{\Omega}/\Omega$, $k^+/\pi^+$, $K^-/\pi^-$, $\bar{p}/\pi^-$, $p/\pi^-$, $\Lambda/\pi^-$, and $\Omega/\pi^-$ which measured in various experimental heavy-ion collisions. The obtained results are covered large energy regions spanning from low energies to high energies such as AGS, SPS and RHIC energies. Also, the HRG calculations of the considered particle ratios are presented in the results for better comparison. Both  the computational ANN model and the phenomenological HRG model have succeeded to describe the suggested particle ratios, as a function of center of mass energy, measured in different high energy heavy-ion collisions. ANN simulations show a highly sensitivity coincidence to the experimental data than HRG model. This conclusion can encourage further use of ANN model in prediction of various particle ratios.

%%%%%%%%%%%%%%%%%%%%%%%%%%%%%%%%%%%%%%%%%%%%%%%%%%%%%%%%%%%%%%%%%%%%%%
%%%   References
%%%%%%%%%%%%%%%%%%%%%%%%%%%%%%%%%%%%%%%%%%%%%%%%%%%%%%%%%%%%%%%%%%%%%%
\bibliographystyle{aip}
\bibliography{Reham_2021}

\end{document}